\documentclass[twocolumn,letterpaper,prb,showpacs,superscriptaddress,amsmath]{revtex4}

\usepackage{graphicx}

\newcommand{\cobaltate}{\mbox{Na$_x$CoO$_2$}}

\begin{document}

\title{The origin of strong correlations and superconductivity in 
Na$_{\mathbf{x}}$CoO$_{\mathbf{2}}$}

\author{Giniyat Khaliullin}
\affiliation{Max-Planck-Institut f\"ur Festk\"orperforschung,
Heisenbergstrasse 1, D-70569 Stuttgart, Germany}

\author{Ji\v{r}\'{\i} Chaloupka}
\affiliation{Max-Planck-Institut f\"ur Festk\"orperforschung,
Heisenbergstrasse 1, D-70569 Stuttgart, Germany}
\affiliation{Department of Condensed Matter Physics, Faculty of Science,
Masaryk University, Kotl\'a\v{r}sk\'a 2, 61137 Brno, Czech Republic}

\begin{abstract}
We propose a minimal model resolving a puzzle of enigmatic correlations
observed in sodium-rich \cobaltate{} where one expects  a simple, free motion
of the dilute $S=1/2$ holes doped into a band insulator NaCoO$_2$. The model
also predicts singlet superconductivity at experimentally observed
compositions. The model is based on a key property of cobalt oxides -- the
spin-state quasidegeneracy of CoO$_6$ octahedral complex -- leading to an
unusual physics of, {\it e.g.}, LaCoO$_3$.  We show that correlated hopping
between $t_{2g}$ and $e_g$ states leads to the spin-polaron physics at 
$x\sim 1$, and to an extended $s$-wave pairing at larger doping when coherent
fermionic bands are formed.    
\end{abstract}

\date{\today}




\maketitle

\section{INTRODUCTION}

Recent studies boosted by the discovery of water-induced superconductivity
(SC) in \cobaltate{} \cite{Tak03} exposed many  remarkable properties of these
compounds \cite{Ong04} such as a spin-sensitive thermopower \cite{Wan03},
unusual charge and spin orderings \cite{Foo04,Ber04,Bay05,Gas06,Ber07}, very
narrow quasiparticle bands \cite{Qia06a,Shi06,Qia06b,Val02,Bro07} {\it etc}. 
While strongly correlated nature of \cobaltate{} is no longer at doubt, the
mechanisms by which the correlated electrons design  such an exotic phase
diagram \cite{Foo04} are not fully understood even on a qualitative level.    

Superconductivity of cobaltates has low $T_c\simeq 5\:\mathrm{K}$. However,
the identification of the pairing mechanism is a problem of principal
importance, because this may shed light on the other puzzles of \cobaltate{}
as well.  Moreover, hopeful comparisons with the high-$T_c$ cuprates have been
made \cite{Tak03,Sch03}, noticing that \cobaltate{} consists of CoO$_2$ layers
with $S=1/2$ Co$^{4+}$ ions doped by $S=0$ Co$^{3+}$ charge carriers, an
apparent $t_{2g}$-band analog of the cuprates.  A triangular lattice formed by
Co ions, providing favorable conditions for a realization of the
resonating-valence-bond (RVB) ideas \cite{And87}, has been also emphasized.  
 
However, it was quickly realized that: ({\it i}) the phase diagram of
\cobaltate{} \cite{Foo04} is radically different from that of cuprates; ({\it
ii}) SC dome is located at valence compositions closer to Co$^{3+}$($S=0$)
rather than Co$^{4+}$($S=1/2$) ({\it i.e.}, at average valences $\lesssim
3.50$) \cite{Tak04,Mil04,Kar04},  not favorable for RVB-theories \cite{Wan04}.
Further, \cobaltate{} shows magnetic order at $x>0.75$ (besides a particular
one at $x=0.5$ \cite{Foo04,Gas06}) which is  counter-intuitive because density
of Co$^{4+}$ spins $\propto (1-x)$ decreases at large $x$.  These observations
make it clear that  the origin and functionality of strong correlations in
cobaltates and cuprates are very different. 

In this paper, we propose a model for strong correlations that operate over
the entire phase diagram of \cobaltate{} and lead to SC optimized near the
valency $3.4$ as observed.  First, we consider a single hole doped in
NaCoO$_2$ and show why its behavior is radically different from that of a free
carrier embedded in a band insulator.  Considering then a Fermi-liquid regime
of \cobaltate{}, we demonstrate how an unusual, kinetic energy driven pairing
emerges in the model. 

The model is based on the following points (none is present in cuprates):
({\it i}) typically, Co$^{3+}$ ions in the octahedral environment possess also
low-lying magnetic states, {\it e.g.} $t_{2g}^5e_g^1$ $S=1$ or $t_{2g}^4e_g^2$
$S=2$; ({\it ii}) in the CoO$_2$ planes with $90^\circ$ \mbox{Co-O-Co} bonds,
the correlated $S=1$ spin states are strongly coupled to the ground state via
the intersite $t_{2g}\leftrightarrow e_g$ hopping (see Fig.~\ref{fig:Hopping})
\cite{note1}.  In other words, the magnetic configuration of Co$^{3+}$ ions is
activated once the mobile Co$^{4+}$ holes are added in NaCoO$_2$. A dynamical
generation of $t_{2g}^5e_g^1$ $S=1$ states by a hole motion converts it into a
many-body correlated object -- the spin-polaron.  At larger density of
Co$^{4+}$, we eliminate a virtual  $S=1$ states perturbatively, and find an
effective model in a form of spin-selective pair hopping of electrons.  The
correlated hopping energy is optimized when holes are paired and condense into
a SC state. 

Spin-state quasidegeneracy  of cobalt ions is well known, LaCoO$_3$ being a
textbook example \cite{Mae04}. According to Ref.~\onlinecite{Hav06}, magnetic
states are in the range of $\sim 200-400\:\mathrm{meV}$ ($S=1$) and $\gtrsim
50\:\mathrm{meV}$ ($S=2$) above the $t_{2g}^6$ $S=0$ ground state (without
lattice relaxations). A balance between the crystal-field, Hund's coupling and
$pd$-covalency is easily tuned and latent magnetism of Co$^{3+}$ living in
virtual states can be activated, {\it e.g.}, by nonmagnetic doping
\cite{Yam96,Cac99}. In oxides with 180$^{\circ}$ $d$-$p$-$d$ bonding as in
LaCoO$_3$, this process leads typically to a ferromagnetic metal stabilized by
an electron promoted into broad $e_g$ bands \cite{Cac99}.  New element of
\cobaltate{} is the 90$^{\circ}$ $d$--$p$--$d$ bonding where the $e_g$--$e_g$
hopping is suppressed. Instead, a large overlap between the neighboring  $e_g$
and $t_{2g}$ orbitals is dominant. A curious situation which arises is that
while Co$^{3+}$ ions are nonmagnetic in NaCoO$_2$,  their $S=1$ $t_{2g}^5e_g$
configurations are dynamically generated in a doped case by the strong
$t_{2g}$--$e_g$ hopping.

A r\'{e}sum\'{e} is that a low-lying magnetic  states of Co$^{3+}$, accessible
for electrons via the intersite hopping,  provide an extra dimension in
physics of \cobaltate{}. In Sec. II, we design a model incorporating this
idea. Based on this model, we demonstrate in Sec. III that a hole doped into
the band insulator NaCoO$_2$ behaves in fact as a magnetic polaron dressed by
the spin-state fluctuations of Co$^{3+}$ ions that are excited by  hole
motion. Sec. IV derives the interaction between holes, mediated by virtual
spin-state excitations of Co$^{3+}$ ions, in a Fermi-liquid regime at finite
hole densities. We also discuss there the relevance of these interactions to
the spin ordering, and find signatures of  $2k_F$-instabilities. Finally, we
focus in Sec. V on the superconductivity and discuss symmetry and doping
dependencies of pairing instabilities within our model. Sec. VI concludes the
paper.       

\section{MODEL HAMILTONIAN}

\begin{figure}[tbp]
\begin{center}
\includegraphics[width=8.3cm]{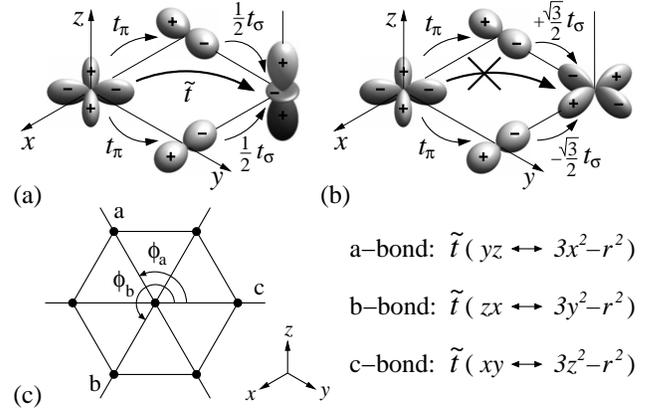}
\caption{
(a)~Electron hopping from $t_{2g}$ to $e_g$ orbital via oxygen atoms in
the case of $90^\circ$ bonds creating $S=1$ $t_{2g}^5e_g$ configuration 
of Co$^{3+}$. The $t_{2g}$ orbital laying in the plane (here
the $xy$ orbital) couples to the out-of plane $e_g$ ($3z^2-r^2$) orbital.
(b)~The coupling to the planar orbital is zero because of the
destructive interference of two channels.  
(c)~Bond directions and corresponding angles in the hexagonal lattice of 
Co ions and $\tilde{t}$-active orbitals on these bonds.
}
\label{fig:Hopping}
\end{center}
\end{figure}

The $t_{2g}$ orbitals in \cobaltate{} split into
$a_{1g}=(d_{xy}+d_{yz}+d_{zx})/\sqrt{3}$ and
$e'_g=(d_{xy}+\mathrm{e}^{\pm\mathrm{i}\varphi}d_{yz}
+\mathrm{e}^{\mp\mathrm{i}\varphi}d_{zx})/\sqrt{3}$ states ($\varphi=2\pi/3$).
The photoemission  experiments \cite{Qia06a,Shi06,Qia06b,Bro07} show that a
single band, derived mostly from the $a_{1g}$ orbitals, is active near the
Fermi level (see Ref.~\onlinecite{Zho05} for the orbital-selection mechanism).
Therefore, we base our model on the $a_{1g}\equiv f$ hole states (its
three-band version will be presented elsewhere \cite{Cha07}). Valence
fluctuations $d_j^6d_i^5\rightarrow d_j^5d_i^6$ within the low-spin $t_{2g}$
manifold read then as 
$H_t=-t\sum_{ij\sigma} f^\dagger_{j\sigma}f^{\phantom{\dagger}}_{i\sigma}$, 
where $t=2t_0/3$ and $t_0=t_{\pi}t_{\pi}/\Delta_{pd}$ is the overlap between
$t_{2g}$ orbitals \cite{Kos03} (hereafter, a hole representation is used).
Our crucial observation is that the $t_{2g}$--$e_g$ hopping $\tilde
t=t_{\sigma}t_{\pi}/\Delta_{pd}$, which uses a stronger $\sigma$-bonding path
with $t_{\sigma}/t_{\pi}\sim 2$, leads to more effective valence fluctuations.
The hopping geometry is depicted in Fig.~\ref{fig:Hopping}.  The nearest
neighbor (NN) Co ions and two O ions binding them determine a plane which
could be labeled $a$, $b$ or $c$ according to the Co-Co bond direction.  With
respect to this plane, the $\tilde{t}$-hopping couples the in-plane $t_{2g}$
orbital to the out-of-plane $e_g$ orbital. 

The $\tilde t$ process generates $S=1$ state of Co$^{3+}$ composed of a
$t_{2g}$ hole and an $e_g$ electron; we represent it by ${\cal T}$ operator
(low-spin $S=0$ $t_{2g}^5e_g^1$ state is much higher in energy and can be
ignored \cite{Cha07}).  ${\cal T}$ is specified by its spin projection and the
$e_g$ orbital $\gamma$ created by $\tilde t$ hopping, {\it i.e.}, 
${\cal T}^\dagger_{+1,\gamma}=
 \overline{e^\dagger_{\gamma\uparrow}f^\dagger_\uparrow}$,
${\cal T}^\dagger_{-1,\gamma}=
 \overline{e^\dagger_{\gamma\downarrow}f^\dagger_\downarrow}$ and
${\cal T}^\dagger_{0,\gamma}=(\overline{
 e^\dagger_{\gamma\uparrow}f^\dagger_\downarrow+
 e^\dagger_{\gamma\downarrow}f^\dagger_\uparrow
})/\sqrt{2}$. 
We are now in position to show our minimal model for \cobaltate{}:
$H_{t-\tilde t}=H_t+H_{\tilde t}$, where  $H_t$ is as given above, while 
\begin{multline}
\label{Htilde}
H_{\tilde t} = -\frac{\tilde t}{\sqrt 3} \sum_{ij} \Bigl[
{\cal T}^\dagger_{+1,\gamma}(i)
f^\dagger_{j\downarrow}f^{\phantom{\dagger}}_{i\uparrow}-
{\cal T}^\dagger_{-1,\gamma}(i)
f^\dagger_{j\uparrow}f^{\phantom{\dagger}}_{i\downarrow}
\\
-{\cal T}^\dagger_{0,\gamma}(i)\,\tfrac1{\sqrt{2}}\!\left(
  f^\dagger_{j\uparrow}f^{\phantom{\dagger}}_{i\uparrow}-
  f^\dagger_{j\downarrow}f^{\phantom{\dagger}}_{i\downarrow}
\right)
+\mathrm{h.c.}\Bigr]\;.
\end{multline}
$H_{\tilde t}$ moves an electron from Co$_j^{3+}$ to Co$_i^{4+}$ -- producing
a $t_{2g}$ hole on site $j$ -- and replaces the $t_{2g}$ hole on site $i$ by a
complex excitation ${\cal T}$. Making use of the   $t_{2g}$--$e_g$ hopping
(the {\it largest} one for 90$^{\circ}$ Co-O-Co bonds), an electron
``picks-up'' the spin correlations in virtual states. The index $\gamma$ is
determined by the orientation of the $\langle ij\rangle$ bond according to the
rules in Fig.~\ref{fig:Hopping}(c). The overlap between $e_g$ orbitals
specified by $\gamma$ and $\gamma'$ is
$\langle\gamma|\gamma'\rangle=\cos(\phi_\gamma-\phi_{\gamma'})$.
Consequently, the excitations ${\cal T}_\gamma$ inherit the same overlap: 
$\langle {\cal T}^{\phantom{\dagger}}_\gamma {\cal T}^\dagger_{\gamma'}\rangle
\propto \langle\gamma|\gamma'\rangle$.
The ${\cal T}$-excitation energy $E_T$ is determined by all the many-body
interactions within the CoO$_6$ complex (Hund's coupling, $p-d$ covalency,
crystal field, \ldots) \cite{Hav06}. This is a free parameter of the model.
Experimentally, $S=1$ states of CoO$_6$ complex in perovskite compound 
LaCoO$_3$ are found at energies $E_T \sim 0.2-0.4\:\mathrm{eV}$ \cite{Hav06} 
as already mentioned in the introduction. 
Based on this observation, we will use in this paper a representative value 
$E_T \simeq 0.3\:\mathrm{eV}$ for layered cobaltates. In units of $a_{1g}$ 
hopping integral $t\simeq 0.1\:\mathrm{eV}$ (which follows from the band
structure fit $t_0\simeq 0.15\:\mathrm{eV}$ \cite{Zho05}), this translates 
into $E_T/t=3$ adopted below in our numerical data. (In principle, we expect
some material dependence of $E_T$ as it is decided by the balance of several
competing interactions. It is therefore highly desirable to quantify a 
multiplet structure of CoO$_6$-complex in \cobaltate{} as done in LaCoO$_3$ 
\cite{Hav06}). For the ratio of the hopping amplitudes $\tilde{t}$ and $t_0$, 
we set $\tilde t/t_0=2$ as $t_{\sigma}/t_{\pi}\sim 2$. 

\section{SPIN-STATE POLARON}

\begin{figure}[btp]
\begin{center}
\includegraphics[width=8.3cm]{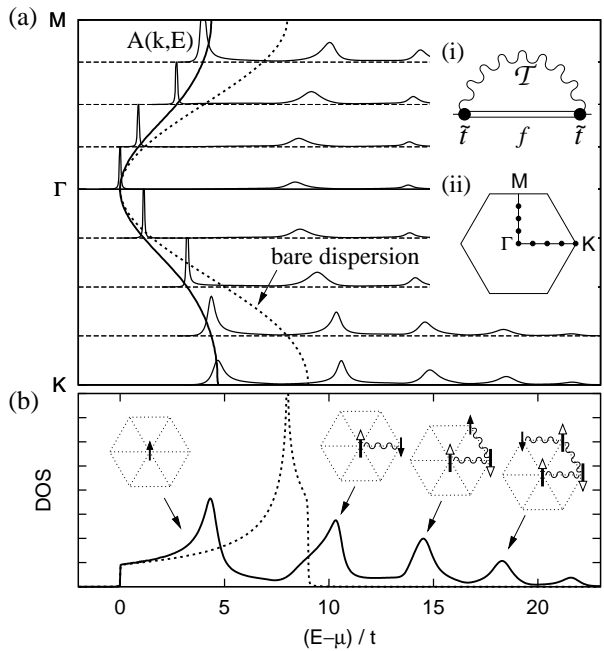}
\caption{
(a)~Spectral functions of a Co$^{4+}$ hole doped in NaCoO$_2$ along 
M-$\Gamma$-K path in the Brillouin zone [inset ({\it ii})]. Bare and 
renormalized dispersions, measured both from the chemical potential 
for better comparison, are shown. (The polaron binding-energy shift 
is $E_b \simeq 2.6t$). ({\it i}) The diagram describing the dressing of a hole 
by $S=1$ $\cal T$-excitations.  
(b)~Density of states compared to that of the bare band. 
Virtual processes associated with the polaronic spectral features 
are sketched: every use of $\tilde{t}$ hopping channel generates 
$S=1$ states of Co$^{3+}$ behind the hole.
}
\label{fig:Spectral}
\end{center}
\end{figure}

It is instructive to consider first a single hole doped in NaCoO$_2$. 
With $H_t$ alone, it is just a usual plane-wave having nothing common with 
what is actually seen in \cobaltate{} at large $x$. Things change radically
when the $H_{\tilde t}$ is switched on: now, a hole generates a multiple of
$\cal T$ excitations [see Fig.~\ref{fig:Spectral}(b)], and spin-polaron
physics of a typical Mott insulator emerges. Given that NaCoO$_2$ itself is
nonmagnetic at all \cite{Lan05}, the correlated behavior of doped holes in
layered cobaltates has been a mystery; it is resolved here by invoking 
a ``virtual Mottness'' of cobalt oxides \cite{Kha05} hidden 
in their low-lying magnetic states. 

Polaron physics is evident from the spectral functions in
Fig.~\ref{fig:Spectral}. We have employed a self-consistent Born
approximation for the selfenergy, which then takes the form:
\begin{equation}
\label{sigma}
\Sigma(\omega)=2\tilde{t}^{\,2}\sum_{\boldsymbol{k}}
\frac{\Gamma_{\boldsymbol{k}}}
{\omega-E_T-\xi_{\boldsymbol{k}}-\Sigma(\omega-E_T)+\mathrm{i}\delta} \;.
\end{equation}
Here, $\Gamma_{\boldsymbol{k}}=c_a^2+c_b^2+c_c^2-c_ac_b-c_bc_c-c_cc_a$ is a
geometrical factor coming from the $e_g$ orbital overlap,
$\xi_{\boldsymbol{k}}=-2t(c_a+c_b+c_c)+\mu$ is the bare dispersion in hole
representation, and $c_\alpha=\cos k_\alpha$ with $k_\alpha$ are projections
of $\boldsymbol{k}$ on $a$, $b$, $c$ axes in the 2D hexagonal lattice of Co
ions [Fig.~\ref{fig:Hopping}(c)].  As the ${\cal T}$-exciton has no dispersion
($e_g$-$e_g$ hopping is zero in $90^{\circ}$-case), the selfenergy is momentum
independent. Strong renormalization of the quasiparticle band and appearance
of the incoherent sidebands as seen in Fig.~\ref{fig:Spectral} are the
characteristic features of polaron formation. Excitations relevant here are
the {\it spin-state} fluctuations of Co$^{3+}$ ions, and a fermionic hole
dressed by these excitations can be termed as {\it spin-state polaron}.
Physically, it is different from a typical magnetic polaron formed in Mott
insulators with a magnetically active ground state, while NaCoO$_2$ is a
nonmagnetic band insulator. 
 
At large $x\sim 1$ limit, dilute polarons are readily trapped by a random
potential of Na-vacancies \cite{Rog07,Mar07}. When the binding is strong,
physics is local and a polaron takes a form of hexagon-shaped $S=1/2$ object
where a hole is oscillating to optimize both $t$ and $\tilde t$ channels.  Our
model provides a microscopic basis for spin-polarons introduced on
experimental grounds \cite{Ber04,Ber07,Bro07} and discussed in  detail in
Refs.~\onlinecite{Kha05,Dag06}. When the density of polarons is increased (as
$x$ decreases), they start to overlap forming narrow bands. Eventually, the
polaron picture breaks down and a correlated Fermi-liquid emerges when $x$ is
further reduced.  

\section{EFFECTIVE INTERACTION BETWEEN $\mathbf{t}_\mathbf{2g}$ HOLES}

\begin{figure}[tbp]
\begin{center}
\includegraphics[width=8.3cm]{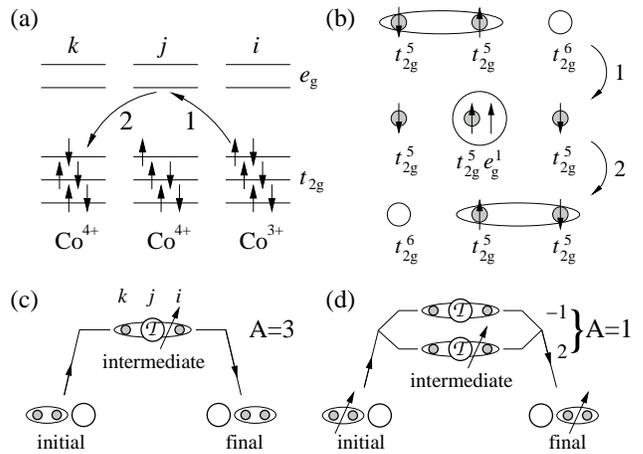}
\caption{
(a)~The physical picture behind Eq.~\eqref{Heff}.  $t_{2g}$ electron of a
$\mathrm{Co}^{3+}_i$ ion moves to  $e_g$ level of the NN $\mathrm{Co}^{4+}_j$
ion (process 1) and then to the $t_{2g}$ level of the next
$\mathrm{Co}^{4+}_k$ neighbor (process 2). This is  depicted in (b) as a
motion of the hole-pair.  (c)~Singlet-pair motion via an intermediate state
composed of $S=1$  $\mathrm{Co}^{3+}_j$ ion ($\cal T$-exciton) and triplet
state of the two holes on $\langle ik\rangle$-bond.  The relative amplitude
$A$ resulting from spin algebra is indicated.  (d)~The triplet-pair hopping
amplitude $A$ is three times smaller because of the destructive interference
of contributions involving singlet and triplet $\langle ik\rangle$-bond.
}
\label{fig:Schematics}
\end{center}
\end{figure}

In the Fermi liquid regime, the Eliashberg-formalism, where the phonon
shake-up processes (triggered by $\cal T$-exciton) can also be incorporated,
would be the best strategy. However, there are delicate constraints to handle:
a lattice site cannot be occupied by two holes or by a hole and ${\cal
T}$-exciton simultaneously.  For the sake of simplicity, we derive an
effective fermionic interaction in a second order perturbation theory in
$\tilde t$ by considering the local virtual process depicted in
Fig.~\ref{fig:Schematics}(a,b). This way, all the constraints in the
intermediate states are treated explicitly. Such a perturbative treatment 
is valid as long as a polaron binding energy $E_b$ (an energy gain due to 
the $\tilde t$ process) is small compared to a bare bandwidth 
$W$ ($\simeq 9t$ in a triangular lattice). From a self-consistent Born
approximation discussed above, we obtained $E_b\simeq 2.6t \sim 0.3W$ for 
$\tilde t=E_T=3t$ used in this paper. For this set of parameters, we can
therefore integrate out a virtual spin states perturbatively.    

As a result, we arrive at the following effective Hamiltonian 
in two equivalent forms:
\begin{align}
\label{Heff}
H_\mathrm{eff}&=\tfrac12 V \sum_{\langle ijk\rangle} \cos(\phi_{ij}-\phi_{jk})
\left[\hat{S}^\dagger_{ij}\hat{S}^{\phantom{\dagger}}_{kj}+
      \tfrac13\hat{\boldsymbol{T}}^\dagger_{ij}
      \hat{\boldsymbol{T}}^{\phantom{\dagger}}_{kj}\right] \\
\label{Heff2}
 &=V\sum_{\langle ijk\rangle} \cos(\phi_{ij}-\phi_{jk}) \left[
n_{j} n_{ik}-\tfrac13 \boldsymbol{s}_{j}\boldsymbol{s}_{ik} \right] \;.
\end{align}
We introduced here a constant $V=\tilde{t}^2/E_T$. Sites $i\neq k$ are the
nearest neighbors of site $j$. The angles $\phi \in (2\pi/3, 4\pi/3, 0)$ are
selected by the orientation of the bonds $\langle ij \rangle$ and $\langle jk
\rangle$ as already explained. No-double-occupancy constraint on $f$ is
implied when using this effective Hamiltonian.  The Hamiltonian in
Eq.~\eqref{Heff} describes the motion of the spin-singlet $\hat{S}_{ij}=
(f_{i\uparrow}f_{j\downarrow}-f_{i\downarrow}f_{j\uparrow})/\sqrt{2}$ and
spin-triplet $\hat{\boldsymbol T}_{ij}= \{f_{i\uparrow}f_{j\uparrow},
\ldots\}$ Co$^{4+}$--Co$^{4+}$ pairs in a background of $S=0$ Co$^{3+}$ ions,
and may lead to the pairing instability as shown below. 

Alternatively, Eq.~\eqref{Heff2} represents the same interaction in a form of
density-density and spin-spin correlations, emphasizing its relevance also to
the charge and spin orderings.  Note that $n_{ik}$ and $\boldsymbol{s}_{ik}$
with $i \neq k$ are the charge and spin densities residing on bonds, 
{\it i.e.}, 
$n_{ik}=\frac12\sum_\sigma
f^\dagger_{i\sigma}f^{\phantom{\dagger}}_{k\sigma}$,
$\boldsymbol{s}_{ik}^z=\frac12\sum_\sigma \sigma
f^\dagger_{i\sigma}f^{\phantom{\dagger}}_{k\sigma}$ 
(while $n_j=n_{jj}$ and $\boldsymbol{s}_j=\boldsymbol{s}_{jj}$ are the usual
on-site operators), so that the interaction acts between the local ({\it
on-site}) and non-local ({\it bond}) operators. In a momentum space,
Eq.~\eqref{Heff2} can be written as 
\begin{equation}
\label{Heff3}
H_\mathrm{eff}= 2V\sum_{\boldsymbol{q}} \left[
n_{-\boldsymbol{q}} \tilde{n}_{\boldsymbol{q}}-\tfrac13 
\boldsymbol{s}_{-\boldsymbol{q}} \tilde{\boldsymbol{s}}_{\boldsymbol{q}}
\right]\;, 
\end{equation}
with the operators $\tilde{n}_{\boldsymbol{q}}=
\tfrac12 \sum_{\boldsymbol{k},\sigma}
F_{\boldsymbol{k}+\boldsymbol{q},\boldsymbol{k}}
f^\dagger_{\boldsymbol{k}+\boldsymbol{q},\sigma}
f^{\phantom{\dagger}}_{\boldsymbol{k},\sigma}$,   
$\tilde{s}^z_{\boldsymbol{q}}=
\tfrac12 \sum_{\boldsymbol{k},\sigma}\sigma 
F_{\boldsymbol{k}+\boldsymbol{q},\boldsymbol{k}}
f^\dagger_{\boldsymbol{k}+\boldsymbol{q},\sigma}
f^{\phantom{\dagger}}_{\boldsymbol{k},\sigma}$, {\it etc.} 
The formfactor $F_{\boldsymbol{k'},\boldsymbol{k}}=
\cos(k_a+k'_a) + \cos(k_b+k'_b) + \cos(k_c+k'_c)
- c_ac'_b - c_bc'_a - c_bc'_c - c_cc'_b - c_cc'_a - c_ac'_c$, 
where $c'_\alpha=\cos k'_\alpha$, originates from a peculiar bond-dependence
of interactions in Eq.~\eqref{Heff2}. It manifests again that the
$\tilde{n}_q$ and $\tilde{\boldsymbol{s}}_q$ operators correspond to the
particle-hole excitations that modulate the charge and spin bonds,
respectively.  
 
\begin{figure}[tbp]
\begin{center}
\includegraphics[width=8.3cm]{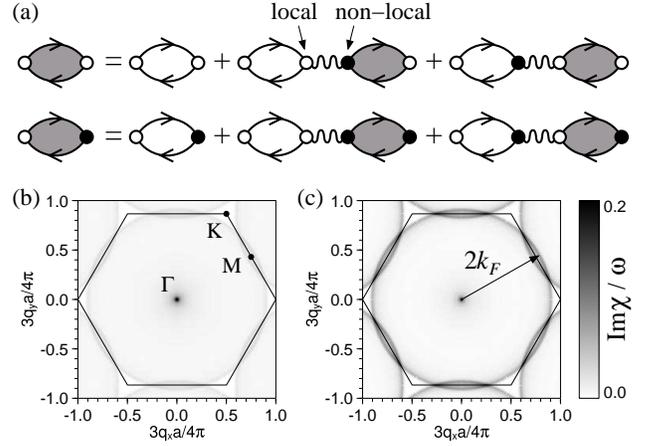}
\caption{
(a)~Diagrammatic representation of RPA equations for the spin susceptibilities
involving the interaction Eq.~\eqref{Heff3} between  $\boldsymbol{s_q}$
(local) and $\tilde{\boldsymbol{s}}_{\boldsymbol{q}}$ (non-local) spin
densities that reside on sites and bonds, respectively.  Bare and RPA-enhanced
susceptibilities are represented by empty and shaded bubbles respectively.
(b)~Map of bare $\chi_s''/\omega$ for $n_d=0.5$ (the average Co-valency $3.5$)
at $T=0.025t$ and $\omega=0.005t$. 
(c)~Corresponding RPA-enhanced susceptibility calculated at
$\tilde{t}^2/E_T=3t$.  The interaction enhances the susceptibility at the
$2k_F$ ring which (at given density $n_d=0.5$) nearly matches the Brillouin
zone boundary.
}
\label{fig:RPA}
\end{center}
\end{figure}

To illustrate this unusual, nonlocal nature of correlations we show in
Fig.~\ref{fig:RPA} the effect of the interaction on the spin susceptibility
within the RPA approximation. The bare spin susceptibility
[Fig.~\ref{fig:RPA}(b)] is concentrated around the $\Gamma$ point. When the
interaction is switched on [Fig.~\ref{fig:RPA}(c)], the $2k_F$ ring in the
susceptibility is enhanced. This suggests the fermionic $2k_F$-instabilities
in a {\it Fermi-liquid} phase, consistent with a picture inferred from the
experiment \cite{Bob06}. Interestingly, the RPA-spin susceptibility at
$n_d=0.5$ is most enhanced near the $M$ point,  {\it i.e.} near the observed
magnetic Bragg peak position \cite{Gas06},   rather than at $K$ typical for
the AF Heisenberg spin system. In order to study the spin ordering at
$n_d=0.5$ more quantitatively, one should take into account also the Na
ordering \cite{Foo04} which breaks a hexagonal symmetry of the underlying
Fermi-surface.

\section{SUPERCONDUCTIVITY DUE TO THE PAIR-HOPPING INTERACTION}

Now, we consider the above Hamiltonian in the context of superconductivity.
It is evident from Eq.~\eqref{Heff} that spin-singlet pairs gain much more
kinetic energy than triplets. As explained in Fig.~\ref{fig:Schematics}, this
nontrivial result originates from a quantum interference between different
realizations of the virtual $\tilde t$  process.  The $S=1$ ${\cal
T}$-excitation in the intermediate state is fully transparent for singlets
which equally use all three $S_z=\pm 1,0$ states of ${\cal T}$.  However, in
the case of triplet pairs, there exist two quantum paths contributing with
opposite signs, which results in a ``spin blockade'' for the motion of
triplets. Alternatively, it can be said that the $S=0$ Co$^{3+}$ states move
more coherently when the $S=1/2$  background is in a singlet state. The
difference from cuprates is that singlets are formed here not due to the
superexchange (in cobaltates, $J$ is small \cite{Wan04}) but because of the
gain in the kinetic energy associated with $\tilde t$ hoppings. 
    
A mean-field BCS analysis of Eq.~\eqref{Heff} shows that $H_\mathrm{eff}$ 
supports either extended $s$-wave singlet SC with the gap function 
$\propto$ $\gamma(\boldsymbol k)=\sqrt{2/3}(c_a +c_b +c_c)$,
or doubly-degenerate spin-triplet $p$-wave pairing with 
$\gamma_{x,y}(\boldsymbol k)=\{(s_a-s_b), (2s_c-s_a-s_b)/\sqrt{3}\}$, 
where $s_{\alpha}=\sin k_{\alpha}$. The $d$-wave channel is repulsive, 
while $f$-wave one is attractive but too weak in the physically reasonable 
doping range. We estimated the $T_c$ from 
\begin{equation}\label{Tceq}
1=\sum_{|\bar{\xi}_{\boldsymbol k}|\le E_T} 
\frac{\bar V_{\alpha}|\gamma_{\alpha}({\boldsymbol k})|^2}
{2\bar\xi_{\boldsymbol k}}\tanh\frac{\bar\xi_{\boldsymbol k}}{2T_c} \;,
\end{equation}
where $\bar V_{\alpha}$ is either $\bar V$ or $\bar V/3$, and the
corresponding formfactors are $\gamma(\boldsymbol k)$ or
$\gamma_{x,y}(\boldsymbol k)$ for the singlet $s$-wave and triplet $p$-wave
pairing, respectively.  To account for the no-double-occupancy constraint, the
fermionic dispersion as well as the pair-hopping amplitude are renormalized by
the Gutzwiller factor \cite{Zho05} $g_t=2n_d/(1+n_d)$ as $(\bar\xi,\bar
V)=(g_t\xi,g_tV)$, where $n_d$ is the relative fraction of Co$^{3+}$ ions.
(The reported Co-valences $\sim$3.4 \cite{Tak04}, $\sim$3.3 \cite{Mil04},
$\sim$3.46 \cite{Kar04} optimal for SC translate then to $n_d=0.6,0.7,0.54$).
In the momentum summation, we have introduced a cutoff equal to the excitation
energy $E_T$. 

\begin{figure}[tbp]
\begin{center}
\includegraphics[width=8.0cm]{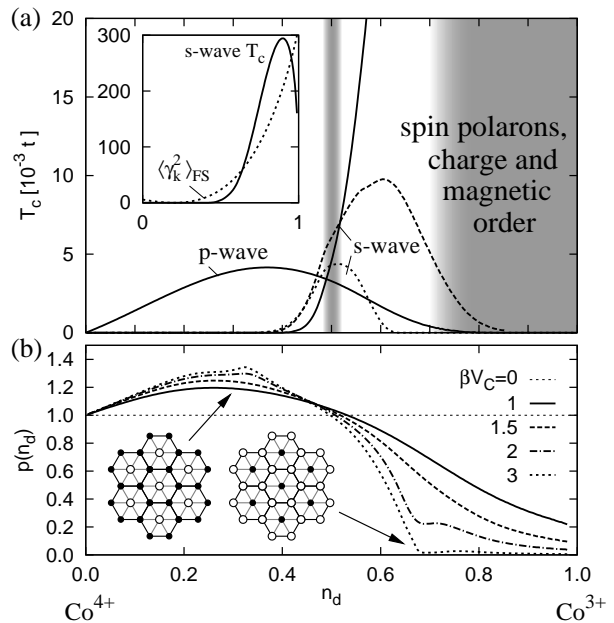}
\caption{
(a)~$T_c$ in the extended $s$-wave and $p$-wave channels.  The complete
profile of the dominant, $s$-wave $T_c$ curve is shown in the left inset
together with $\gamma_{\boldsymbol{k}}^2$ (in arbitrary units) on the Fermi
surface.  The dashed ($\beta V_C=1.5$) and dotted ($\beta V_C=3$) $T_c$ curves
are calculated including NN Coulomb repulsion which reduces the pairing
interaction $V$ at large $n_d$.  Shaded regions indicate the observed
competing orderings (including the spin-charge order at $n_d=0.5$). 
(b)~Probability ratio $p(n_d)$ (see text for definition) renormalizing the
pairing interaction at different values of NN Coulomb repulsion relative to
the effective temperature $1/\beta \propto$ bandwidth.  The feature at
$n_d=1/3$ for large $V_C$ manifests a honeycomb-lattice formation where each
Co$^{3+}$ ($\circ$) has the maximum possible number of neighboring
Co$^{4+}$--Co$^{4+}$ pairs ($\bullet$--$\bullet$). Above $n_d=2/3$, Co$^{4+}$
holes can avoid each-other completely if $V_C$ is sufficiently large.
}
\label{fig:Tc}
\end{center}
\end{figure}

We solved Eq.~\eqref{Tceq} at $V=3t$ (as it follows from $\tilde t=E_T=3t$
used in previous sections). In terms of the BCS-coupling constant, 
this translates into $\lambda=\bar V\bar N=VN \sim 1/3$ considering 
the density of states $N\sim 1/W\sim 1/9t$. Therefore, the present 
formulation in terms of an effective fermionic Hamiltonian (3) 
should give a reasonable results. At larger values
of $V$, we encounter a strong coupling regime where one should use
instead the original model (1) and treat a virtual spin states explicitely.  
This limit remains a challenging problem for future study. 

The resulting $T_c$ values from Eq.~\eqref{Tceq} are presented 
in Fig.~\ref{fig:Tc}(a) as solid lines. As expected, 
the highest $T_c$ values are found in the singlet channel,
increasing with Co$^{3+}$ density due to the formfactor effect, until SC
disappears at $n_d=1$ limit.  A weak triplet pairing is present thanks to its
formfactor matching well the Fermi surface, but it is expected to be destroyed
by ({\it e.g.} Na) disorder.  (We should notice that these trends are based on
the present mean-field decoupling which ignores a collective spin
fluctuations. One can speculate, for instance, that the triplet pairing may be
supported by a ferromagnetic fluctuations within the CoO$_2$ planes observed
\cite{Bay05} at {\it large} $n_d$ limit). 

As the SC pairing considered here is due to the pair-hopping, Coulomb
repulsion between the holes will oppose it. This is not a big trouble at high
density of Co$^{4+}$ spins (as they cannot avoid themselves) but becomes a
severe issue in a spin-diluted regime at large $n_d$, where Coulomb repulsion
reduces the process described in Fig.~\ref{fig:Schematics} hence the amplitude
$V$.  Instead, the formation of spatially separated spin-polarons
(Fig.~\ref{fig:Spectral}) is favored, and competing orderings take over, such
as an in-plane ferromagnetism induced by a residual interactions between
spin-polarons \cite{Dag06}.  To include the effects related to the Coulomb
repulsion in the Gutzwiller fashion, we use an additional multiplicative
factor reflecting the suppression of the probability $P_{ijk}$ of having the
required Co$^{3+}_i-$Co$^{4+}_j-$Co$^{4+}_k$ configuration.  We have
determined this probability using a classical Monte-Carlo simulation of
hardcore particles with NN Coulomb repulsion $V_C$.  The simulations were
performed at different ``effective temperatures'' $1/\beta$ imitating the
kinetic energy (of the order of bandwidth) which competes with the Coulomb
repulsion in the real system.  Plotted in Fig.~\ref{fig:Tc}(b) is the
probability ratio $p(n_d)=P_{ijk}(V_C)/P_{ijk}(V_C=0)$ for several values of
$\beta V_C$.  The corresponding $T_c$ curves calculated with 
$\bar V\rightarrow p(n_d)\bar V$ locate the SC-dome near the valence $3.4$, 
in a remarkable correspondence with experiment \cite{Tak04,Mil04,Kar04}.     

Finally, our $t-\tilde t$ model provides a clear hint on the role of
water-intercalation needed for SC in \cobaltate{}. Without water, a random
Na-potential induces some amount of spin-polarons locally (the origin of
``Curie-Weiss metal'' \cite{Foo04}) which suppress the pairing among the
remaining fermions the usual way. Once this potential is screened-out by the
water layers, an intrinsic ground state of CoO$_2$ planes as in
Fig.~\ref{fig:Tc} is revealed.  (This interpretation of the water effect is
consistent with the absence of superconductivity in the monolayer hydrate of
Na$_x$CoO$_2$, where the water resides in the Na layers.) The remaining
``enemy'' of SC is the Coulomb repulsion which prevents the pairing of dilute
Co$^{4+}$ fermions and supports the formation of spin-polarons and magnetism
instead. More pronounced polaron physics (because of the presence of large
$S=1$ $\cal T$-exciton and narrow bandwidth) explains why $T_c$ in cobaltates
is low compared to cuprates.

\begin{figure}[tbp]
\begin{center}
\includegraphics{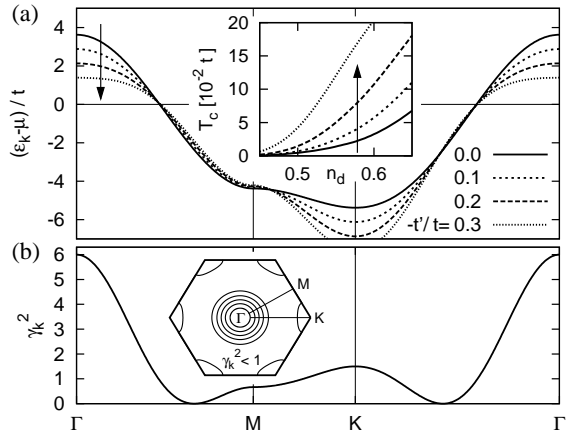}
\caption{
(a)~Band structure for different values of $t'/t$ ($n_d=0.6$) and the
corresponding effect on the $s$-wave transition temperature. As the band
dispersion near $\Gamma$ point comes closer to the Fermi level, it can exploit
the larger formfactor and $T_c$ increases as shown in the inset. The
increase is also partially induced by the decreased Fermi velocity along
$\Gamma$-M direction.
(b)~Formfactor of the extended $s$-wave pairing interaction. 
The inset shows the contours for $\gamma_{\boldsymbol k}^2=1,2,3,4,5$. 
}
\label{fig:Bands}
\end{center}
\end{figure}

Another mechanism for the water effect is provided by the band-structure
calculations\cite{Joh04} that indicate a substantial flattening of the
$a_{1g}$ band-top and a reduction of the band splitting when the water-layers
are present. To study the former effect, we include negative $t'$ in our
calculation. Due to the combined effect of better formfactor utilization in
$s$-wave channel and Fermi velocity reduction this enhances singlet pairing as
presented in Fig.~\ref{fig:Bands}.  In triplet channel, on the other hand, the
$t'$-effect is weaker as it includes only the latter factor, {\it i.e.}, the
enhancement due to the reduced Fermi velocity ($p$-wave formfactor utilization
is not very sensitive to $t'$).

\section{CONCLUSIONS}

To summarize, we have presented $t-\tilde t$ model for \cobaltate{} 
which is based on the spin-state quasidegeneracy of CoO$_6$ octahedral 
complex in oxides and the specific lattice geometry of the 
CoO$_2$ planes in layered cobaltates. The model naturally explains 
the strong correlations found in the sodium-rich region
due to the spin-polaron formation. We derived effective interactions in a
Fermi-liquid regime and discussed their impact on spin fluctuations.  The
model predicts superconductivity mediated by the spin-state fluctuations of
Co$^{3+}$ ions, at experimentally observed compositions.  The basic idea
behind the model is that  due to the 90$^{\circ}$ \mbox{$d$-$p$-$d$} pathway
in the edge-shared structure, the electron transport in \cobaltate{} is
entangled with low-lying $S=1$ magnetic states of Co$^{3+}$ ions which become
an essential part of the \cobaltate{} physics.  Given the simplicity and
experimentally motivated design of the model, its success can hardly be
accidental. Therefore, $t-\tilde t$ Hamiltonian can be regarded as a basic
minimal model for \cobaltate{}. It may also have broader applications, {\it
e.g.}, in oxides of Rh and Ir ions with a similar spin-orbital structure and
lattice geometry. 

\section*{ACKNOWLEDGMENTS}

We would like to thank B.~Keimer, J.~Sirker, and M.Z.~Hasan for stimulating 
discussions. 
This work was partially supported by the Ministry of Education of Czech
Republic (MSM0021622410).


\begin{thebibliography}{99}
\bibitem{note1}
$t_{2g}^4e_g^2$ $S=2$ configuration is not accessible by hopping.
\end{thebibliography}
\end{document}